\documentclass[aps,pre,twocolumn]{revtex4-1}
\usepackage{amssymb}
\usepackage{amsmath,bm}
\usepackage{graphicx,color}
\usepackage{bbm}
\usepackage{multirow}
\newcommand{\B}[1]{{\bm{#1}}}

\begin{document}

\title{Universal Scaling Laws for Shear Induced Dilation in Frictional Granular Media}

\author{M. M. Bandi$^1$, Prasenjit Das$^2$, Oleg Gendelman$^3$, H. George E. Hentschel$^2$ and Itamar Procaccia$^2$ }
\affiliation{$^1$ Nonlinear and Non-equilibrium Physics Unit, OIST Graduate University, Okinawa, Japan 904-0495, $^2$Department of Chemical Physics, The Weizmann Institute of Science, Rehovot 76100, Israel, $^3$ Faculty of Mechanical Engineering, Technion, Haifa 32000, Israel.}

\begin{abstract}
Compressed frictional granular matter cannot flow without dilation. Upon forced shearing to generate flow, the amount of dilation may depend on the initial preparation and a host of material variables. On the basis of both experiments and numerical simulations we show that as a result of training by repeated compression-decompression cycles the amount of dilation induced by shearing the system depends only on the shear rate and on the (pre-shearing) packing fraction. Relating the rheological response to structural properties allows us to derive a scaling law for the amount of dilation after $n$ cycles of compression-decompression. The resulting scaling law has a universal exponent that for trained systems is independent of the inter-granules force laws, friction parameters and strain rate. The amplitude of the scaling law is analytically computable, and it depends only on the shear rate and the asymptotic packing fraction.
\end{abstract}
\maketitle
\section{Introduction}
Compressed granular media, with or without friction, are jammed, and cannot flow without
dilation \cite{66Bag,96JNB,02Lem,11Sir}. Subjected to shear rate by external forces, such media dilate, reducing the packing fraction in regions that participate in flows. The dilation may be very inhomogeneous, and may depend on a host of parameters that characterize the granular assembly. Understanding the resulting rheology is complicated due to the inherent properties of granular matter, like frictional losses, arching, segregation and thixotropy \cite{91TG}. These complications result in a paucity of universal results, and the literature of frictional granular rheology at finite strain rate offers a bewildering array of particular examples that are not easy to comprehend, resisting attempts to organize and systematize \cite{84Sav,95TH,02AS,10LS}.
In this paper we do not attempt to add to the existing literature on dilation {\em per se}.
There exist an extensive list of papers that provide very useful studies of the phenomenon, see for example Refs.~\cite{04MiDi,05daCruz,06JFP,08FP,08PR,10DD}. Our aim here is to discuss the effects of aging that results from cyclic training of the system {\em before} shear rate is applied. Apparently this aspect of the problem had not been considered before.

In recent studies it became apparent that some universal results can be gleaned by training
the system under repeated cycles of compression-decompression \cite{10MTB,13RDB,13BRKE}, building a memory that ``cleans" the system from random effects present in ``as compressed" frictional granular systems. For example it was shown that the packing fraction converges under repeated cycles to an asymptotic value
following a universal law \cite{18BHPRZ}. Another example is the universal giant friction slip event that occurs when the pressure goes to zero upon unjamming \cite{18HPR}. Here we follow on this line of reasoning and study the dilation induced by shear rate after training the system by $n$ compression-decompression cycles. Indeed we find an enormous simplification resulting in a universal power law that characterizes the amount of dilation observed after training with $n$ cycles. The power law indicates that training and memory result in the amount of dilation becoming a function of the strain rate and the packing fraction only. (Here we refer to the packing fraction {\em before} shearing is applied). The exponent of the scaling law is independent of the working pressure, the strain rate, the friction parameters and the force laws between granules.

The paper is organized as follows: in Sect.~\ref{training} we introduce the main ideas needed
for the present study; we review briefly some recent results on training
and memory formation. In Sect.~\ref{simulations} we describe the numerical simulations
of dilation under shear, and present the theory and the numerical data for the universal
scaling law for the dilation after $n$ cycles of training. In Sect.~\ref{experiments} we
describe the experiments that provide further support for the universality of
the main power law discovered in this paper. Sect.~\ref{conclusions} present a summary and some concluding remarks.

\section{training by compression-decompression cycles}
\label{training}
\begin{figure}
\includegraphics[scale=0.80]{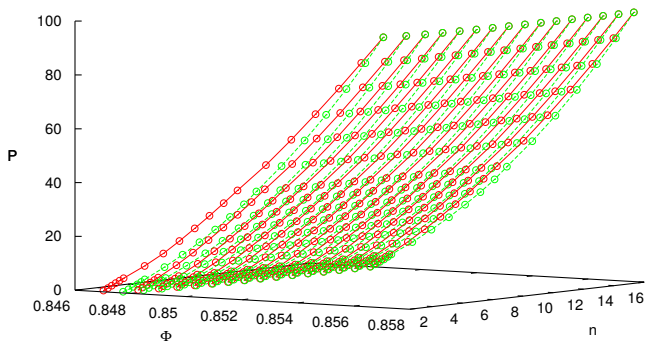}
\caption{(Color online). Typical hysteresis loops obtained numerically upon uniaxial $n$ cycles ofcompression and decompression of an amorphous configuration of frictional disks. Here an assembly of disks with binary sizes are compressed from above in the presence of gravity. The pressure $P$ is on the upper piston, measured in dimensionless units of $mg/d$, see text for details.  The packing fraction $\Phi$ is dimensionless. Compression legs are in red and decompression in green. The $n$ axis is
provided for clarity, since the hysteresis curves are more compressed
than in cyclic training without gravity, see \cite{18BHPRZ}.}
\label{numloops}
\end{figure}

When frictional granular media are trained by cyclic loading and
unloading \cite{13BRKE,13RLR,14FFS, 18BHPRZ} memory is introduced in the system.  Here we refer to training by uniaxial compression until the pressure reaches a maximal value $P_{\rm max}$ after which the the system is decompressed back to zero pressure, and then compressed again. In our
experiments and numerical simulations compression and decompression are achieved by one moving wall and ``pressure" always refer to the external pressure on this moving wall. In each cycle the packing fraction is increased until it reaches an asymptotic limit. During compression and decompression dissipation leads to hysteresis, but with repeated cycles the dissipation diminishes to a finite limit and the system retains memory of an asymptotic loaded state that is not forgotten even under complete unloading. An example of such training protocol as observed in numerical simulations \cite{18BHPRZ} is shown in Fig.~\ref{numloops}. Similar experimental training
protocols were described for example in Ref.~\cite{18BHPRZ}.

Associated with the reduced dissipation and the increase in memory one finds a universal power law in the packing fraction $\Phi_n$ after the $n$th cycle. This scalings is  expected to hold irrespective of the details of the microscopic interactions. In every compression leg of the cycle the system compactifies, until a limit $\Phi$ value is reached for the chosen maximal pressure.  To quantify this process we can measure the volume fraction $\Phi_n(P_{\rm max})$ at the highest value of the pressure in the $n$th cycle. Define then a new variable
\begin{equation}
X_n \equiv \Phi_{n+1}(P_{\rm max})-\Phi_n(P_{\rm max}) \ .
\end{equation}
This new variable is history dependent in the sense that $X_{n+1} = g(X_n)$ where the function $g(x)$ is unknown at this
point. This function must have a fixed point $g(x=0) =0$ since the series $\sum_n X_n$ must converge; for any given
chosen maximal pressure there is a limit volume fraction that cannot be exceeded. Near the fixed point, assuming analyticity, we expect the form
\begin{equation}
X_{n+1} = g(X_n) = X_n -C X_n^2 + \cdots \ .
\end{equation}
The solution of this equation for $n$ large is
\begin{equation}
X_n = \frac{C^{-1}}{n} \ . \label{scaling}
\end{equation}
A direct measurement of $X_n$ as a function of $n$ in the present simulations which are
recorded below is
shown in the log-log plot presented in Fig.~\ref{Xnvsn}. In Ref.~\cite{18BHPRZ} one can find arguments and evidence for the generality of this power law and for the existence of a fixed-point asymptotic reversible loop.

\section{Dilation under shear: simulations}
\label{simulations}
\subsection{preparation}

The granular system that we simulate consists of disks of mass $m=1$ and diameters $d=1$ and $1.4d$ in equal numbers.
To prepare the system for shear and dilation we begin with a ``box" of fixed horizonal length (in the $x$ direction) of 60$d$ and a height (in the $y$ direction) of 160$d$. To start, 2000 small and 2000 large disks are placed randomly without overlaps. The upper wall has a mass $M=100$ that is free to move; gravity is chosen such that $g=1$.  The moving upper wall and the fixed lower wall are made of particles of identical properties and diameters in the continuous range of $[d,2d]$.  Applying periodic boundary condition in the horizontal direction we now
apply a small pressure $P$ on the upper wall. We simulate the system using molecular dynamics with Hertz normal forces and Mindlin tangential forces as described below. We solve Newton's equations of motion with linear damping in the velocities of the disks. For a given pressure the simulation
continues until mechanical equilibrium is reached. The pressure is then increased in small steps
followed by equilibration until the desired final pressure is obtained.
An example of an initial configuration is shown in Fig.~\ref{incon}.
\begin{figure}
\includegraphics[scale=0.25]{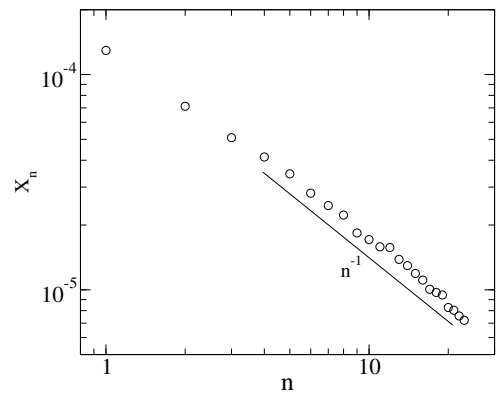}
\caption{Log-log plot of $X_n\equiv \Phi_{n+1}(P_{\rm max})-\Phi_n(P_{\rm max})$ vs. $n$. The black dots are the data, the solid line is the theoretical inverse power law prediction. The data corroborates Eq.~(\ref{scaling}).}
\label{Xnvsn}
\end{figure}

\begin{figure}
\includegraphics[scale=0.75]{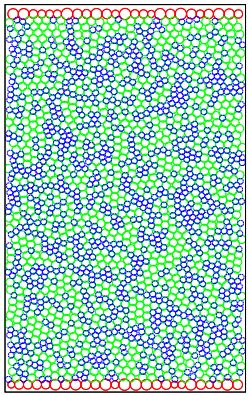}
\caption{(Color online). An example of a typical initial configuration in the numerical simulation. The blue disks are the large ones and the green the small ones. Red disks are glued to the walls.}
\label{incon}
\end{figure}

The contact fores (both the normal and tangential forces which arises due to friction) are modeled according to the DEM (discrete element method) developed by Cundall and Strack \cite{79CS}. Implementation of static friction is done via tracking the elastic part of the shear displacement from the time contact was first formed.
When the disks are compressed they interact via both
normal and tangential forces.  Particles $i$ and $j$, at positions ${\B r_i, \B r_j}$ with velocities ${\B v_i, \B v_j}$ and angular velocities ${\B \omega_i, \B \omega_j}$ will  experience a relative normal compression on contact given by $\Delta_{ij}=|\B r_{ij}-D_{ij}|$, where $\B r_{ij}$ is the vector joining the centers of mass and $D_{ij}=R_i+R_j$; this gives rise to a  normal force $ \B F^{(n)}_{ij} $. The normal force is modeled as a Hertzian contact, whereas the tangential force is given by a Mindlin force \cite{79CS}. Defining $R_{ij}^{-1}\equiv R_i^{-1}+R_j^{-1}$, the force magnitudes are,
\begin{eqnarray}
\B F^{(n)}_{ij}\!&=&\!k_n\Delta_{ij} \B n_{ij}-\frac{\gamma_n}{2} \B {v}_{n_{ij}}\ , \:
\B F^{(t)}_{ij}\!=\!-k_t \B t_{ij}-\frac{\gamma_t}{2} \B {v}_{t_{ij}} \\
k_n &=& k_n^{'}\sqrt{ \Delta_{ij} R_{ij}} \ , \quad
k_t = k_t^{'} \sqrt{ \Delta_{ij} R_{ij}} \\
\gamma_{n} &=& \gamma_{n}^{'}  \sqrt{ \Delta_{ij} R_{ij}}\ , \quad
\gamma_{t} = \gamma_{t}^{'}  \sqrt{ \Delta_{ij} R_{ij}} \ . \label{forces}
\end{eqnarray}
Here $\Delta _{ij}$ and $t_{ij}$ are normal and tangential displacement; $\B n_{ij}$ is the normal unit vector.  $k_n^{'}=2\times 10^5$ and $k_t^{'}=2k_n'/7$ are spring stiffness for normal and tangential mode of deformation: $\gamma_n^{'}=50$ and $\gamma_t^{'}=50$ are viscoelastic damping constant for normal and tangential deformation.
   $\B {v}_{n_{ij}}$ and $\B {v}_{t_{ij}}$ are respectively normal and tangential component of the relative velocity between two particles. The relative normal and tangential velocity are given by
   \begin{eqnarray}
\B {v}_{n_{ij}}&=& (\B {v}_{ij} .\B n_{ij})\B n_{ij}  \\
\B {v}_{t_{ij}}&=& \B {v}_{ij}-\B {v}_{n_{ij}} - \frac{1}{2}(\B \omega_i + \B \omega_j)\times \B r_{ij} \ . \label{velocities}
\end{eqnarray}
   where $\B {v}_{ij} = \B {v}_{i} - \B {v}_{j}$. Elastic tangential displacement $ \B t_{ij}$ is set to zero when the contact is first made and is calculated using $\frac{d \B t_{ij}}{d t}= \B {v}_{t_{ij}}$ and also the rigid body rotation around the contact point is accounted for to ensure that $ \B t_{ij}$ always remains in the local tangent plane of the contact \cite{01SEGHLP}.

The translational and rotational acceleration of particles are calculated from Newton's second law; total forces and torques on particle $i$ are given by
\begin{eqnarray}
\B F^{(tot)}_{i}&=& \sum_{j}\B F^{(n)}_{ij} + \B F^{(t)}_{ij}  \\
\B \tau ^{(tot)}_{i}&=& -\frac{1}{2}\sum_{j}\B r_{ij} \times \B F^{(t)}_{ij}.
\end{eqnarray}
The tangential force varies linearly with the relative tangential displacement at the contact point as long as the tangential
   force does not exceed the Coulomb limit
   \begin{equation}
   F^{(t)}_{ij} \le \mu F^{(n)}_{ij} \ , \label{Coulomb}
   \end{equation}
  where $\mu$ is a material dependent coefficient. When this limit is exceeded the contact slips in a dissipative
  fashion. In our simulations we keep the
  magnitude of $t_{ij}$  so that $F^{(t)}_{ij} =\mu F^{(n)}_{ij}$. The direction of $\B t_{ij}$
  is allowed to change if further slip takes place.
\subsection{Shearing and Dilating}

\begin{figure}
\includegraphics[scale=0.25]{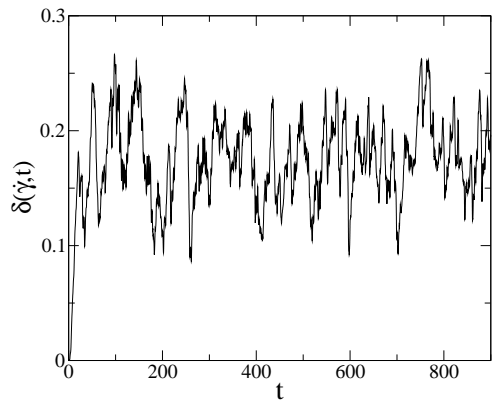}
\caption{A typical trajectory of $\delta(t)$ after 9 training cycles. Here $P=5$, $\dot\gamma L_y(0)= 0.1$ and the friction coefficient $\mu=0.1$.}
\label{deltaoft}
\end{figure}

Having compacted the granular medium through a certain number of cycles, we next examine what happens if this same medium is subjected to a shear strain at a rate $\dot \gamma$ on its upper surface. Flow is possible only by dilating the material especially close to the upper moving wall \cite{84Sav,95TH,02AS,10LS}. Denoting the rest height of the box by $L_y(0)$ we measure the actual height of the upper wall which is a function of time and the shear rate, denoted as $L_y(t,\dot \gamma)$. The dilation is now denoted by $\delta(t,\dot \gamma)$ where
\begin{equation}
\delta(t,\dot \gamma) \equiv L_y(t) -L_y(0) \ .
\end{equation}
The time dependence of $\delta(t)$ is quite complex. A typical trajectory of this quantity is
shown in Fig.~\ref{deltaoft}.
\begin{figure}
\includegraphics[scale=0.25]{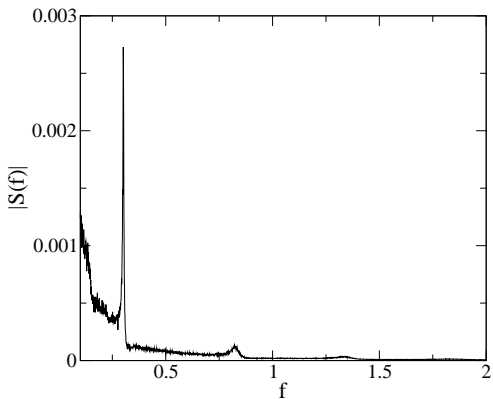}
\caption{A Fourier transform of the trajectory shown in Fig.~\ref{deltaoft}. }
\label{spectrum}
\end{figure}
Obviously, the trajectory indicates some noisy periodicity around some average. To extract the dominant frequency of the response of the upper wall we can compute the
Fourier transform of this trajectory,
\begin{eqnarray}
S(f) &\equiv& \frac{1}{500}\int_{t=200}^{700} dt~ [\delta(t,\dot \gamma) -\langle \delta \rangle (\dot \gamma)]~e^{i2\pi f t}\ ,\\
\langle \delta \rangle (\dot \gamma) &\equiv& \frac{1}{500}\int_{t=200}^{700} dt~ \delta(t,\dot \gamma) \ .
\end{eqnarray}
The limits of integration were chosen to eliminate the initial rise to a `steady state' and to ensure convergence of the result. Averaging such spectra over 50 independent initial configurations results in a typical spectrum as seen in Fig.~\ref{spectrum}.
The spectrum
is dominated by one typical frequency. The nature of this frequency and its dependence on
shear rate are interesting by themselves, but they fall outside the scope of the present
paper. We only note in passing that the principal frequency (the main peak
in Fig.~\ref{spectrum}) is fully understandable as a result of the excitation of the primary
bulk elastic mode of the system.

\begin{figure}
	\vskip 1.8cm
\includegraphics[scale=0.40]{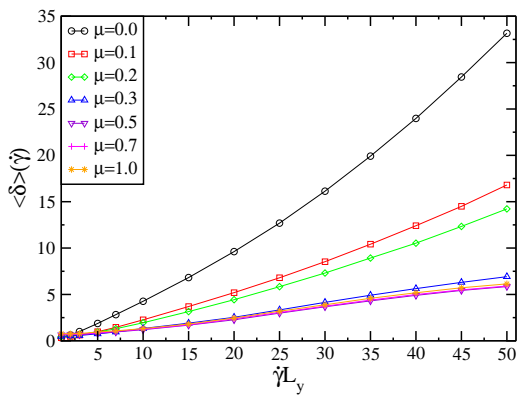}
\caption{(Color online). The dependence of the average dilation on shear rate and friction coefficient. This
data was obtained with 2000 particles as in the experiment below. }
\label{dependence}
\end{figure}

Having computed the average dilation $\langle \delta \rangle(\dot\gamma)$ we can next examine its
dependence on the parameters of the model. The average dilation obviously depends on the
many variables, including the packing fraction, the shear rate, the friction coefficient
etc. For {\em a given packing fraction} we can study the dependence on the shear rate and
the friction coefficient. Typical results are presented in Fig.~\ref{dependence}.
One sees that the average dilation increases with shear rate. In fact the dependence of the dilation on the shear rate when $\dot\gamma\to 0$ appears to be non-analytic. The discussion of
this non-analyticity is interesting but will be defer to a future publication. We also note that the amount of dilation reduces with the friction coefficient, leading to convergence in the plot
for $\mu>0.3$.

The question that remains is how can we obtain data collapse and universal statements about
the dilation under shear. To this aim we need to develop a bit of theory.

\subsection{Data Collapse and Universal Laws}
\subsubsection{Cyclic Training}

Motivated by the universal scaling laws for the packing fraction as described in
Sect.~\ref{training} and Fig.~\ref{Xnvsn} we study next the physics of dilation
in cyclically trained systems.
The cyclic training is achieved by uniaxial straining such that the pressure is increased by pushing down the upper wall in quasistatic fashion until we reach a maximal chosen pressure; in
the present simulations this pressure is
$P_{\rm max}=100$. After each compression step, the system is allowed to relax to reach a new mechanical equilibrium. After a full compression leg, a cycle is completed by decompressing back to zero pressure, where the next compression cycle begins. The packing fraction $\Phi$ is monitored throughout this process. Each such cycle traces a hysteresis loop in the $P-\Phi$ plane, see Fig.~\ref{numloops} as an example.

The measurements of average dilation will be made now after $n-1$ cycles. The system is
decompressed to zero at the $n-1$'th cycle, and then compressed again to a chosen value
of the pressure $P_{\rm w}$. At that pressure we then strain the system at a given strain rate
$\dot \gamma$ to measure the average dilation. To get better statistics we repeat
the whole procedure to obtain $\langle \delta_n \rangle(\dot \gamma) $ averaged over many
realizations.

To achieve universal results it is always prudent to work with dimensionless quantities.
Thus instead of working with $\langle \delta_n \rangle(\dot \gamma) $  we opt to define
a new, related quantity which is dimensionless. To define this dimensionless quantity
denote by $\Phi_n(P_{\rm w})$ the packing fraction associated with the unstrained systems in
the $n$th cycle. After settling into the steady state with a give shear rate $\dot\gamma$
the asymptotic average packing fraction is denoted as $\Phi^*_n \left(\Phi_n(P_{\rm w}),\dot\gamma\right)$. The dimensionless dilation is then
\begin{equation}
D(\Phi_n(P_{\rm w}), \dot \gamma)\equiv \big[ \frac{\Phi_n(P_{\rm w})}{\Phi_n^*(\Phi_n(P_{\rm w}), \dot \gamma)} - 1 \big]
\ .
\label{avdil}
\end{equation}
Needless to say, besides being dimensionless the dependence of this measure on the shear rate
and on the friction coefficient remains identical to the data shown in Fig.~\ref{dependence}.
To simplify the notation we use below $D_n(\dot\gamma)\equiv D(\Phi_n(P_{\rm w}), \dot \gamma)$.

\subsubsection{Universal scaling law}

Having at our disposal the universal scaling law for the series $X_n$ it is natural to consider the series of differences in dimensionless dilations $D_{n+1}-D_n$. The main result of this
subsection will be the series of these differences can be re-scaled to become (for large $n$) independent
of $\dot\gamma$, the initial pressure, the friction coefficient etc. To see how to achieve
this simplification we note that after many cycles, when $\Phi_n\to \Phi_\infty$, we can write

\begin{equation}
D_{n+1}( \dot \gamma)-D_n( \dot \gamma) \approx D'(\Phi_{\infty}, \dot \gamma) X_n \approx  \frac{D'(\Phi_{\infty}, \dot \gamma)/C}{n} .
\end{equation}
where $D'(\Phi_{\infty}, \dot \gamma)= dD(\Phi, \dot \gamma)/d\Phi |_{\Phi = \Phi_{\infty}}$.
Besides the immediate consequence that dilation difference tends to zero as $1/n$, we also predict that the amplitude of this scaling appears to be a universal coefficient
$D'(\Phi_{\infty}, \dot \gamma)/C$. Since $C$ is known from the data on the packing fraction itself, we need here to examine the coefficient $D'(\Phi_{\infty}, \dot \gamma)$.

\begin{figure}
\includegraphics[scale=0.25]{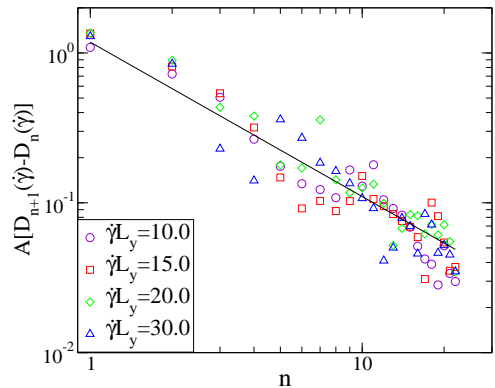}
\caption{(Color online). $A[D_{n+1}-D_n]$ vs $n$. The line is a best fit reading $1.18 n^{-1.03}$.
The data shown are for $\mu=0.1$ but for other values of $\mu$ the re-scaled data
collapse on the same curve. }
\label{final}
\end{figure}

To compute the coefficient we start from Eq.~(\ref{avdil}) and write
\begin{equation}
\label{Ddef3}
\Phi_n D'(\Phi , \dot \gamma) = [1+ D(\Phi_n , \dot \gamma)]- [1+ D(\Phi_n , \dot \gamma)]^2 d\Phi_n^*/d\Phi_n .
\end{equation}
Now if we assume that as $n\rightarrow \infty$ the granular medium looses its memory of its initial condition then we would expect that $d\Phi_n^*/d\Phi_n \rightarrow 0$ and asymptotically
we will find
\begin{equation}
\label{Ddef4}
D'(\Phi_{\infty} , \dot \gamma) \approx \frac{[1+ D(\Phi_{\infty} , \dot \gamma)]}{\Phi_{\infty}}  .
\end{equation}
and the asymptotic scaling of the dilation can be written as
\begin{equation}
D_{n+1}( \dot \gamma)-D_n( \dot \gamma) \approx  \frac{[1+ D(\Phi_{\infty} , \dot \gamma)]/(\Phi_{\infty}C)}{n} .
\end{equation}
Finally, denoting
\begin{equation}
A^{-1}\equiv [1+ D(\Phi_{\infty} , \dot \gamma)]/(\Phi_{\infty}C)\ ,
\label{A}
\end{equation}
we expect that $A[D_{n+1}-D_n]$ should become independent of any parameter in the problem
yielding a universal power law $1/n$.
This prediction is tested against the numerical simulations and the results are shown
in Fig.~\ref{final}. We find that instead of unity the pre-factor is close to 1.2.
Of course constants of the order of unity are permissible in this theory.

\section{Dilation under shear: experiments}
\label{experiments}
\subsection{Setup Description}
\begin{figure}
\includegraphics[scale=0.24]{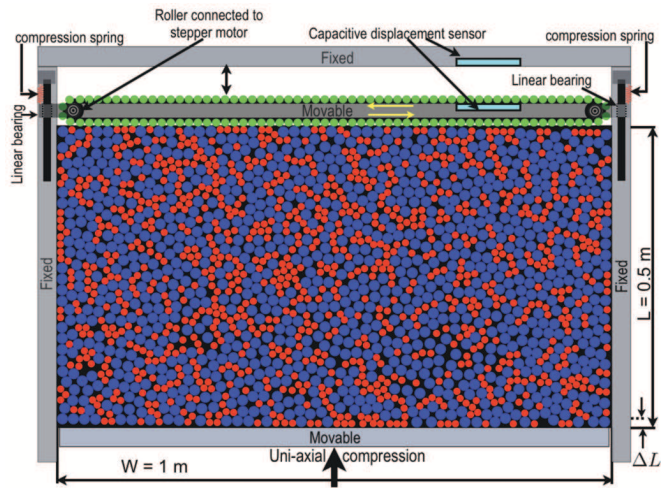}
\caption{(Color online) Schematic of the granular shear dilation experimental setup.}
\label{fig1}
\end{figure}

The experimental setup (see fig.~\ref{fig1}) is a variant of the uni-axial compression setup presented in Ref. \cite{18BHPRZ}. It is comprised of a quasi-2D rectangular cell of length $L = 0.5$m and width $W = 1$m filled with a bidispersed set of of photoelastic disks (diameter 1 cm and 1.5 cm) in equal proportion. The disks were prepared in-house by curing liquid polymer where the modulus and friction coefficient could be tuned independently; details of disk preparation are presented in Ref.~\cite{Akella2018}.

The two boundary walls separated by width ($W$) were held fixed while the opposing boundaries separated by length ($L$) were movable. The bottom movable wall shown in the schematic fig.~\ref{fig1} was used to achieve uni-axial compression from one end while the opposing wall was used to shear the system. At any given time, only one of these two walls was movable while holding the opposing wall fixed. The bottom wall employed in uni-axial compression was controlled by a motorized translation stage with 500 nm precision using home-built capacitive displacement sensors; full implementation details and translation protocols are given in supplemental information of Ref.~\cite{18BHPRZ}. During the initial uni-axial compression and decompression cycles, the upper shear boundary had its linear bearings clamped rigid by electromagnetic actuators.
After completion of the uni-axial compression and decompression cycles for training the granular configuration, the bottom uni-axial compression boundary was clamped rigid and the upper shear boundary had its electromagnetic clamps turned off. The shear boundary consisted of an aluminum plate with two rollers at each end (please see fig.~\ref{fig1}) with a belt drive running along the width ($W$) and looping around the rollers. Disks of small diameter (1 cm) were stuck to the belt (green disks in schematic fig.~\ref{fig1}). Both rollers were connected by individual belt drives to a stepper motor that was placed separately, i.e. decoupled from the setup, so that vibrations from the stepper did not perturb the granular setup. We note an experimental design trade off in using stepper motors as against servo motors. Whereas DC servo motors provide excellent speed (shear rate) control characteristics, they stall when the load increases on the belt drive due to dilation. Stepper motors on the other hand handle very high torques but can only operate at low speeds (shear rates). This design constraint necessitating the use to stepper motors limits the experimental range of shear rates below that achieved in the numerical simulations reported here.

When the electromagnetic clamps were not actuated, the shear boundary could move uni-axially by means of two high lubrication linear bearings at either end of the boundary. Two soft compression springs were placed between the rigid (Fixed) frame and the movable aluminum plate holding the shear belt drive to provide nominal uni-axial compressive force to the shear boundary. This was necessary so that the shear boundary would not decouple completely and move away from the granular pack due to dilation during shear. Soft springs provide the nominal compressive force of 2.6 N sufficient to keep the shear boundary in contact with the granular pack, but being soft they do not impact dilation within measurement tolerance of dilation. The displacement due to dilation was measured using a home-built parallel place capacitive displacement sensor mounted on the fixed and movable shear boundaries as shown in the schematic (fig.~\ref{fig1}). Stray field from the linear bearings' electromagnetic clamps introduced noise in the capacitive displacement sensor, for this reason the precision in dilation displacement is of order 1 micrometer. Be that as it may, since the maximum dilation displacement observed in our measurements never exceeded 0.5 cm, perhaps owing to low shear rates, assured that the 1 micrometer displacement precision was comfortable.

Before closing description of the experimental setup, we point that our experimental design is not ideal to study shear. The simulations employed in this study apply periodic boundary conditions along the horizontal, i.e. shear axis direction. In our experiments, the two fixed boundaries separated by width $W$ give rise to spurious effects not observed in simulations. Ideally such an experiment ought to involve a Couette style annular geometry as implemented in Ref.~\cite{Miller1996, Veje1999} among other works. We have adopted the design detailed above since it affords a quick adaptation to our uni-axial compression setup. Despite these design shortcomings and the systematics introduced by non-periodic or rigid boundaries, as we show, the experiments do capture the primary features of the numerical results, especially the predicted power-law scaling by theory.

\subsection{Experimental Protocol}
The setup was prepared by placing the disks in the experimental chamber in a random configuration and then the movable shear boundary was electromagnetically clamped rigid. The bottom compression boundary was then activated to train the pack through repeated uni-axial compression and decompression cycles in quasi-static steps of 1 micrometer displacement, the step-size of the motorized translation stage being controlled {\it via} feedback sensing from the capacitive displacement sensor as explained in Ref.~\cite{18BHPRZ}. The compression was performed in each cycle up to a maximum two-dimensional pressure $P_{max} =  300$ N/m and a total of $n = 25$ compression-decompression cycles were performed to train the configuration. At the end of $n = 25$ cycles, the bottom uni-axial compression boundary was clamped rigid and the electromagnetic clamps of the movable shear boundary were deactivated. The stepper motor speed was then set to the desired shear rate, the initial distance between the top-fixed and movable shear boundary was recorded with the capacitive sensor and the shear was turned on. The dilation due to shear was continuously monitored through the change in displacement recorded by the capacitive displacement sensor which was sampled at 10 KHz. Following an initial transient, the dilation settled to a steady value modulo instantaneous fluctuations from shear of the configuration. A long-time average over 100 seconds of shear was then taken to obtain the average dilation displacement.
\begin{figure}
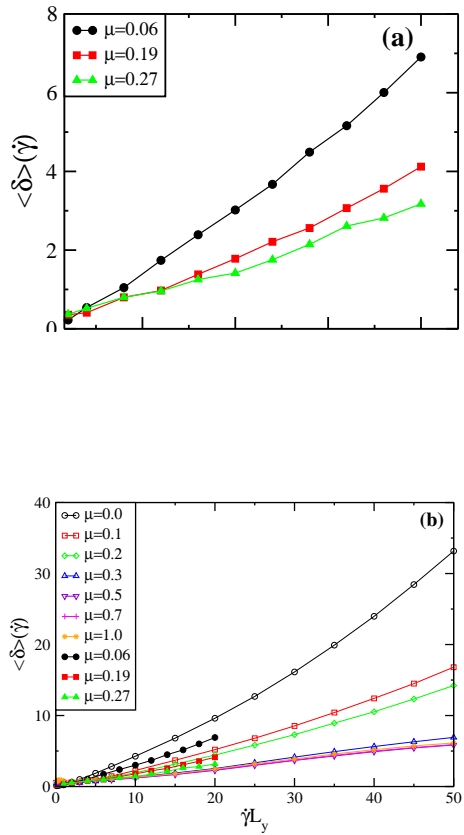

\includegraphics[scale=0.485]{dilationFig9a.eps}
\vskip 1.3 cm
\includegraphics[scale=0.35]{dilationFig9b.eps}
\caption{(Color online). The dependence of the average dilation on shear rate and friction coefficient. Upper panel: experimental results for three values of the friction coefficient. Lower panel: experimental and simulational results shown together. Note how well the data for many values of the friction
coefficient enmesh properly, showing a close correspondence between the simulations and the experiment.}
\label{exptdelta}
\end{figure}

\subsection{Experimental Results}
To present the experimental results we begin with the dilation as a function
of shear rate. In the upper panel of Fig.~\ref{exptdelta} we present the experimental data. In
 the lower panel the experimental data is overlayed
on the simulation data that was shown before in Fig.~\ref{dependence}.

\begin{figure}
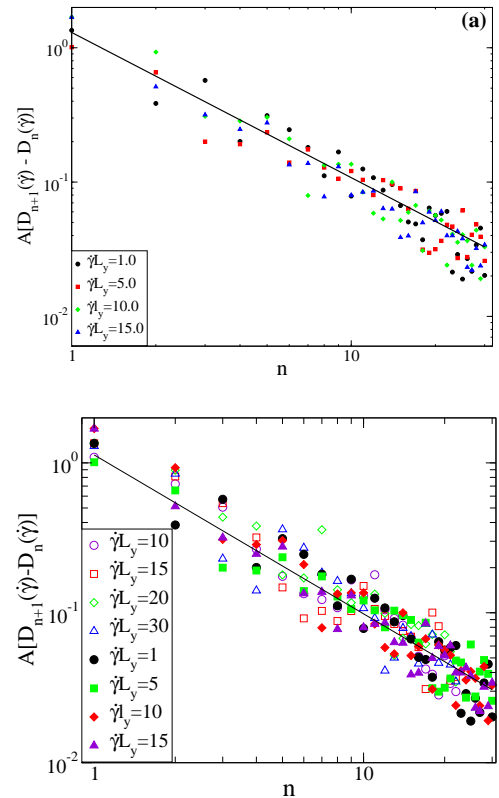

\vskip 0.3 cm
\includegraphics[scale=0.24]{dilationFig10a.eps}
\vskip 0.5 cm
\includegraphics[scale=0.25]{dilationFig10b.eps}
\caption{(Color online). Upper panel: The experimental results for $A[D_{n+1}-D_n]$ vs $n$ for four different shear rates. The black line is a best fit reading $1.3n^{-1.08}$.
Lower panel: the experimental results together with the simulation results, with the black line being the best fit reading $1.13 n^{-1.06}$.}
\label{expfinal}
\end{figure}

One observes that the data of experiments and simulations mesh together quite well,
despite differences in boundary condition that are discussed in the next subsection.
The experimental analog of Fig.~\ref{final} is shown in the upper panel of Fig.~\ref{expfinal}.
Our conclusion is that the experimental results lend a strong support to the theory
and to simulation results.
\subsection{Comparison of theory and experiments}
There are protocol differences between the simulations and experiments. Firstly, the simulations employ compression and shear using the same boundary whereas the experiments apply compression-decompression cycles from one boundary and apply shear from the opposite boundary. Secondly, the simulations employ periodic boundary conditions which the experiments do not implement as explained in the previous section. Frictional granular media are well known for strong dependence on protocol and preparation history. Despite the differences in protocol between simulations and experiments, we find the $1/n$ power-law scaling holds in the experiments pointing to the robustness of the theoretical predictions. Admittedly, the constant pre-factor in the experiments is slightly higher than that observed in simulations but it is still $\mathcal{O}(1)$. Similarly, the experimental power-law slope is also slightly greater than $1/n$ which we attribute to systematic effects arising from fixed boundaries, but the power-law trend does exist.
\section{Concluding Remarks}
\label{conclusions}
The main point of this paper is that training a frictional granular system by compression-decompression cycles can ``clean" an ``as compressed" system from
random effects that complicate the interpretation of rheological properties.
In the present example we examined the amount of dilation caused by shearing
the system with a give shear rate $\dot\gamma$. After training with $n$ compression-
decompression cycles the scaled dilation $D_n(\dot\gamma)$ could be predicted
since the series converges with a universal scaling exponent $n^{-1}$. The pre-factors
could be also estimated from the knowledge of the equally generic $n^{-1}$ dependence
of the associated series $\Phi_{n}$ of the packing fraction after $n$ cycles. One should
note that in both the simulations and in the experiments finite size effects are significant, introducing errors of
the order of unity in the coefficient $C$ of Eq.~(\ref{scaling}), in
 the value of
$\Phi_\infty$ and in the values of $D(\Phi_{\infty} , \dot \gamma)$. All these enter the
coefficient $A$ of Eq.~(\ref{A}). Together they contribute to the scatter seen in
Figs.~\ref{final} and \ref{expfinal}. Taking all this into account we consider the agreement between
theory and measurements quite satisfactory. It appears quite worthwhile to continue
in the future to examine the effects of training and memory on disordered systems and their
rheology.

\acknowledgments
This theoretical work has been supported in part by the US-Israel Binational Science Foundation,
the Israel Science Foundation under the Israel-Singapore program, and the Joint Laboratory
with the Universita' di Roma "La Sapienza".
The experiments were supported by the Nonlinear and Non-equilibrium Physics Unit, OIST Graduate University. MMB gratefully acknowledges generous hosting by The International Centre for Theoretical Sciences, Bengaluru, India under the auspices of the ``Entropy, Information and Order in Soft Matter'' program while working on this manuscript

\bibliography{ALL}

\begin{thebibliography}{27}%
\makeatletter
\providecommand \@ifxundefined [1]{%
 \@ifx{#1\undefined}
}%
\providecommand \@ifnum [1]{%
 \ifnum #1\expandafter \@firstoftwo
 \else \expandafter \@secondoftwo
 \fi
}%
\providecommand \@ifx [1]{%
 \ifx #1\expandafter \@firstoftwo
 \else \expandafter \@secondoftwo
 \fi
}%
\providecommand \natexlab [1]{#1}%
\providecommand \enquote  [1]{``#1''}%
\providecommand \bibnamefont  [1]{#1}%
\providecommand \bibfnamefont [1]{#1}%
\providecommand \citenamefont [1]{#1}%
\providecommand \href@noop [0]{\@secondoftwo}%
\providecommand \href [0]{\begingroup \@sanitize@url \@href}%
\providecommand \@href[1]{\@@startlink{#1}\@@href}%
\providecommand \@@href[1]{\endgroup#1\@@endlink}%
\providecommand \@sanitize@url [0]{\catcode `\\12\catcode `\$12\catcode
  `\&12\catcode `\#12\catcode `\^12\catcode `\_12\catcode `\%12\relax}%
\providecommand \@@startlink[1]{}%
\providecommand \@@endlink[0]{}%
\providecommand \url  [0]{\begingroup\@sanitize@url \@url }%
\providecommand \@url [1]{\endgroup\@href {#1}{\urlprefix }}%
\providecommand \urlprefix  [0]{URL }%
\providecommand \Eprint [0]{\href }%
\providecommand \doibase [0]{http://dx.doi.org/}%
\providecommand \selectlanguage [0]{\@gobble}%
\providecommand \bibinfo  [0]{\@secondoftwo}%
\providecommand \bibfield  [0]{\@secondoftwo}%
\providecommand \translation [1]{[#1]}%
\providecommand \BibitemOpen [0]{}%
\providecommand \bibitemStop [0]{}%
\providecommand \bibitemNoStop [0]{.\EOS\space}%
\providecommand \EOS [0]{\spacefactor3000\relax}%
\providecommand \BibitemShut  [1]{\csname bibitem#1\endcsname}%
\let\auto@bib@innerbib\@empty
\bibitem [{\citenamefont {Bagnold}(1966)}]{66Bag}%
  \BibitemOpen
  \bibfield  {author} {\bibinfo {author} {\bibfnamefont {R.~A.}\ \bibnamefont
  {Bagnold}},\ }\href {\doibase 10.1098/rspa.1966.0236} {\bibfield  {journal}
  {\bibinfo  {journal} {Proc. Roy. Soc. A}\ }\textbf {\bibinfo {volume}
  {295}},\ \bibinfo {pages} {219} (\bibinfo {year} {1966})}\BibitemShut
  {NoStop}%
\bibitem [{\citenamefont {Jaeger}\ \emph {et~al.}(1996)\citenamefont {Jaeger},
  \citenamefont {Nagel},\ and\ \citenamefont {Behringer}}]{96JNB}%
  \BibitemOpen
  \bibfield  {author} {\bibinfo {author} {\bibfnamefont {H.~M.}\ \bibnamefont
  {Jaeger}}, \bibinfo {author} {\bibfnamefont {S.~R.}\ \bibnamefont {Nagel}}, \
  and\ \bibinfo {author} {\bibfnamefont {R.~P.}\ \bibnamefont {Behringer}},\
  }\href {\doibase 10.1103/RevModPhys.68.1259} {\bibfield  {journal} {\bibinfo
  {journal} {Rev. Mod. Phys.}\ }\textbf {\bibinfo {volume} {68}},\ \bibinfo
  {pages} {1259} (\bibinfo {year} {1996})}\BibitemShut {NoStop}%
\bibitem [{\citenamefont {Lema\^{\i}tre}(2002)}]{02Lem}%
  \BibitemOpen
  \bibfield  {author} {\bibinfo {author} {\bibfnamefont {A.}~\bibnamefont
  {Lema\^{\i}tre}},\ }\href {\doibase 10.1103/PhysRevLett.89.195503} {\bibfield
   {journal} {\bibinfo  {journal} {Phys. Rev. Lett.}\ }\textbf {\bibinfo
  {volume} {89}},\ \bibinfo {pages} {195503} (\bibinfo {year}
  {2002})}\BibitemShut {NoStop}%
\bibitem [{\citenamefont {Sirono}(2011)}]{11Sir}%
  \BibitemOpen
  \bibfield  {author} {\bibinfo {author} {\bibfnamefont {S.}~\bibnamefont
  {Sirono}},\ }\href {http://stacks.iop.org/0295-5075/96/i=3/a=30003}
  {\bibfield  {journal} {\bibinfo  {journal} {EPL (Europhysics Letters)}\
  }\textbf {\bibinfo {volume} {96}},\ \bibinfo {pages} {30003} (\bibinfo {year}
  {2011})}\BibitemShut {NoStop}%
\bibitem [{\citenamefont {Thompson}\ and\ \citenamefont {Grest}(1991)}]{91TG}%
  \BibitemOpen
  \bibfield  {author} {\bibinfo {author} {\bibfnamefont {P.~A.}\ \bibnamefont
  {Thompson}}\ and\ \bibinfo {author} {\bibfnamefont {G.~S.}\ \bibnamefont
  {Grest}},\ }\href@noop {} {\bibfield  {journal} {\bibinfo  {journal} {Phys.
  Rev. Lett.}\ }\textbf {\bibinfo {volume} {67}},\ \bibinfo {pages} {1751}
  (\bibinfo {year} {1991})}\BibitemShut {NoStop}%
\bibitem [{\citenamefont {Savage}(1984)}]{84Sav}%
  \BibitemOpen
  \bibfield  {author} {\bibinfo {author} {\bibfnamefont {S.~B.}\ \bibnamefont
  {Savage}},\ }\href {\doibase https://doi.org/10.1016/S0065-2156(08)70047-4}
  {\bibfield  {journal} {\bibinfo  {journal} {Advances in Applied Mechanics}\
  }\textbf {\bibinfo {volume} {24}},\ \bibinfo {pages} {289 } (\bibinfo {year}
  {1984})}\BibitemShut {NoStop}%
\bibitem [{\citenamefont {Tillemans}\ and\ \citenamefont
  {Herrmann}(1995)}]{95TH}%
  \BibitemOpen
  \bibfield  {author} {\bibinfo {author} {\bibfnamefont {H.-J.}\ \bibnamefont
  {Tillemans}}\ and\ \bibinfo {author} {\bibfnamefont {H.~J.}\ \bibnamefont
  {Herrmann}},\ }\href {\doibase https://doi.org/10.1016/0378-4371(95)00111-J}
  {\bibfield  {journal} {\bibinfo  {journal} {Physica A: Statistical Mechanics
  and its Applications}\ }\textbf {\bibinfo {volume} {217}},\ \bibinfo {pages}
  {261 } (\bibinfo {year} {1995})}\BibitemShut {NoStop}%
\bibitem [{\citenamefont {Aharonov}\ and\ \citenamefont {Sparks}(2002)}]{02AS}%
  \BibitemOpen
  \bibfield  {author} {\bibinfo {author} {\bibfnamefont {E.}~\bibnamefont
  {Aharonov}}\ and\ \bibinfo {author} {\bibfnamefont {D.}~\bibnamefont
  {Sparks}},\ }\href {\doibase 10.1103/PhysRevE.65.051302} {\bibfield
  {journal} {\bibinfo  {journal} {Phys. Rev. E}\ }\textbf {\bibinfo {volume}
  {65}},\ \bibinfo {pages} {051302} (\bibinfo {year} {2002})}\BibitemShut
  {NoStop}%
\bibitem [{\citenamefont {Lőrincz}\ and\ \citenamefont {Schall}(2010)}]{10LS}%
  \BibitemOpen
  \bibfield  {author} {\bibinfo {author} {\bibfnamefont {K.~A.}\ \bibnamefont
  {Lőrincz}}\ and\ \bibinfo {author} {\bibfnamefont {P.}~\bibnamefont
  {Schall}},\ }\href {\doibase 10.1039/B926817K} {\bibfield  {journal}
  {\bibinfo  {journal} {Soft Matter}\ }\textbf {\bibinfo {volume} {6}},\
  \bibinfo {pages} {3044} (\bibinfo {year} {2010})}\BibitemShut {NoStop}%
\bibitem [{\citenamefont {{GDR MiDi}}(2004)}]{04MiDi}%
  \BibitemOpen
  \bibfield  {author} {\bibinfo {author} {\bibnamefont {{GDR MiDi}}},\ }\href
  {\doibase 10.1140/epje/i2003-10153-0} {\bibfield  {journal} {\bibinfo
  {journal} {The European Physical Journal E}\ }\textbf {\bibinfo {volume}
  {14}},\ \bibinfo {pages} {341} (\bibinfo {year} {2004})}\BibitemShut
  {NoStop}%
\bibitem [{\citenamefont {da~Cruz}\ \emph {et~al.}(2005)\citenamefont
  {da~Cruz}, \citenamefont {Emam}, \citenamefont {Prochnow}, \citenamefont
  {Roux},\ and\ \citenamefont {Chevoir}}]{05daCruz}%
  \BibitemOpen
  \bibfield  {author} {\bibinfo {author} {\bibfnamefont {F.}~\bibnamefont
  {da~Cruz}}, \bibinfo {author} {\bibfnamefont {S.}~\bibnamefont {Emam}},
  \bibinfo {author} {\bibfnamefont {M.}~\bibnamefont {Prochnow}}, \bibinfo
  {author} {\bibfnamefont {J.-N.}\ \bibnamefont {Roux}}, \ and\ \bibinfo
  {author} {\bibfnamefont {F.~m.~c.}\ \bibnamefont {Chevoir}},\ }\href
  {\doibase 10.1103/PhysRevE.72.021309} {\bibfield  {journal} {\bibinfo
  {journal} {Phys. Rev. E}\ }\textbf {\bibinfo {volume} {72}},\ \bibinfo
  {pages} {021309} (\bibinfo {year} {2005})}\BibitemShut {NoStop}%
\bibitem [{\citenamefont {Jop}\ \emph {et~al.}(2006)\citenamefont {Jop},
  \citenamefont {Forterre},\ and\ \citenamefont {Pouliquen}}]{06JFP}%
  \BibitemOpen
  \bibfield  {author} {\bibinfo {author} {\bibfnamefont {P.}~\bibnamefont
  {Jop}}, \bibinfo {author} {\bibfnamefont {Y.}~\bibnamefont {Forterre}}, \
  and\ \bibinfo {author} {\bibfnamefont {O.}~\bibnamefont {Pouliquen}},\
  }\href@noop {} {\bibfield  {journal} {\bibinfo  {journal} {Nature}\ }\textbf
  {\bibinfo {volume} {441}},\ \bibinfo {pages} {727–7301} (\bibinfo {year}
  {2006})}\BibitemShut {NoStop}%
\bibitem [{\citenamefont {Forterre}\ and\ \citenamefont
  {Pouliquen}(2008)}]{08FP}%
  \BibitemOpen
  \bibfield  {author} {\bibinfo {author} {\bibfnamefont {Y.}~\bibnamefont
  {Forterre}}\ and\ \bibinfo {author} {\bibfnamefont {O.}~\bibnamefont
  {Pouliquen}},\ }\href@noop {} {\bibfield  {journal} {\bibinfo  {journal}
  {Annual Review of Fluid Mechanics}\ }\textbf {\bibinfo {volume} {40}},\
  \bibinfo {pages} {1} (\bibinfo {year} {2008})}\BibitemShut {NoStop}%
\bibitem [{\citenamefont {Peyneau}\ and\ \citenamefont {Roux}(2008)}]{08PR}%
  \BibitemOpen
  \bibfield  {author} {\bibinfo {author} {\bibfnamefont {P.-E.}\ \bibnamefont
  {Peyneau}}\ and\ \bibinfo {author} {\bibfnamefont {J.-N.}\ \bibnamefont
  {Roux}},\ }\href {\doibase 10.1103/PhysRevE.78.011307} {\bibfield  {journal}
  {\bibinfo  {journal} {Phys. Rev. E}\ }\textbf {\bibinfo {volume} {78}},\
  \bibinfo {pages} {011307} (\bibinfo {year} {2008})}\BibitemShut {NoStop}%
\bibitem [{\citenamefont {Drozd}\ and\ \citenamefont {Denniston}(2010)}]{10DD}%
  \BibitemOpen
  \bibfield  {author} {\bibinfo {author} {\bibfnamefont {J.~J.}\ \bibnamefont
  {Drozd}}\ and\ \bibinfo {author} {\bibfnamefont {C.}~\bibnamefont
  {Denniston}},\ }\href {\doibase 10.1103/PhysRevE.81.021305} {\bibfield
  {journal} {\bibinfo  {journal} {Phys. Rev. E}\ }\textbf {\bibinfo {volume}
  {81}},\ \bibinfo {pages} {021305} (\bibinfo {year} {2010})}\BibitemShut
  {NoStop}%
\bibitem [{\citenamefont {Zhang}\ \emph {et~al.}(2010)\citenamefont {Zhang},
  \citenamefont {Majmudar}, \citenamefont {Tordesillas},\ and\ \citenamefont
  {Behringer}}]{10MTB}%
  \BibitemOpen
  \bibfield  {author} {\bibinfo {author} {\bibfnamefont {J.}~\bibnamefont
  {Zhang}}, \bibinfo {author} {\bibfnamefont {T.~S.}\ \bibnamefont {Majmudar}},
  \bibinfo {author} {\bibfnamefont {A.}~\bibnamefont {Tordesillas}}, \ and\
  \bibinfo {author} {\bibfnamefont {R.~P.}\ \bibnamefont {Behringer}},\
  }\href@noop {} {\bibfield  {journal} {\bibinfo  {journal} {Granular Matter}\
  }\textbf {\bibinfo {volume} {12}},\ \bibinfo {pages} {159} (\bibinfo {year}
  {2010})}\BibitemShut {NoStop}%
\bibitem [{\citenamefont {Ren}\ \emph {et~al.}(2013)\citenamefont {Ren},
  \citenamefont {Dijksman},\ and\ \citenamefont {Behringer}}]{13RDB}%
  \BibitemOpen
  \bibfield  {author} {\bibinfo {author} {\bibfnamefont {J.}~\bibnamefont
  {Ren}}, \bibinfo {author} {\bibfnamefont {J.~A.}\ \bibnamefont {Dijksman}}, \
  and\ \bibinfo {author} {\bibfnamefont {R.~P.}\ \bibnamefont {Behringer}},\
  }\href {\doibase 10.1103/PhysRevLett.110.018302} {\bibfield  {journal}
  {\bibinfo  {journal} {Phys. Rev. Lett.}\ }\textbf {\bibinfo {volume} {110}},\
  \bibinfo {pages} {018302} (\bibinfo {year} {2013})}\BibitemShut {NoStop}%
\bibitem [{\citenamefont {Bandi}\ \emph {et~al.}(2013)\citenamefont {Bandi},
  \citenamefont {Rivera}, \citenamefont {Krzakala},\ and\ \citenamefont
  {Ecke}}]{13BRKE}%
  \BibitemOpen
  \bibfield  {author} {\bibinfo {author} {\bibfnamefont {M.~M.}\ \bibnamefont
  {Bandi}}, \bibinfo {author} {\bibfnamefont {M.~K.}\ \bibnamefont {Rivera}},
  \bibinfo {author} {\bibfnamefont {F.}~\bibnamefont {Krzakala}}, \ and\
  \bibinfo {author} {\bibfnamefont {R.~E.}\ \bibnamefont {Ecke}},\ }\href
  {\doibase 10.1103/PhysRevE.87.042205} {\bibfield  {journal} {\bibinfo
  {journal} {Phys. Rev. E}\ }\textbf {\bibinfo {volume} {87}},\ \bibinfo
  {pages} {042205} (\bibinfo {year} {2013})}\BibitemShut {NoStop}%
\bibitem [{\citenamefont {Bandi}\ \emph {et~al.}(2018)\citenamefont {Bandi},
  \citenamefont {Hentschel}, \citenamefont {Procaccia}, \citenamefont {Roy},\
  and\ \citenamefont {Zylberg}}]{18BHPRZ}%
  \BibitemOpen
  \bibfield  {author} {\bibinfo {author} {\bibfnamefont {M.~M.}\ \bibnamefont
  {Bandi}}, \bibinfo {author} {\bibfnamefont {H.~G.~E.}\ \bibnamefont
  {Hentschel}}, \bibinfo {author} {\bibfnamefont {I.}~\bibnamefont
  {Procaccia}}, \bibinfo {author} {\bibfnamefont {S.}~\bibnamefont {Roy}}, \
  and\ \bibinfo {author} {\bibfnamefont {J.}~\bibnamefont {Zylberg}},\ }\href
  {http://stacks.iop.org/0295-5075/122/i=3/a=38003} {\bibfield  {journal}
  {\bibinfo  {journal} {Europhys. Lett.}\ }\textbf {\bibinfo {volume} {122}},\
  \bibinfo {pages} {38003} (\bibinfo {year} {2018})}\BibitemShut {NoStop}%
\bibitem [{\citenamefont {Hentschel}\ \emph {et~al.}()\citenamefont
  {Hentschel}, \citenamefont {Procaccia},\ and\ \citenamefont {Roy}}]{18HPR}%
  \BibitemOpen
  \bibfield  {author} {\bibinfo {author} {\bibfnamefont {H.~G.~E.}\
  \bibnamefont {Hentschel}}, \bibinfo {author} {\bibfnamefont {I.}~\bibnamefont
  {Procaccia}}, \ and\ \bibinfo {author} {\bibfnamefont {S.}~\bibnamefont
  {Roy}},\ }\href@noop {} {\bibinfo  {journal} {arXiv: 1805.11439v1}\
  }\BibitemShut {NoStop}%
\bibitem [{\citenamefont {Regev}\ \emph {et~al.}(2013)\citenamefont {Regev},
  \citenamefont {Lookman},\ and\ \citenamefont {Reichhardt}}]{13RLR}%
  \BibitemOpen
\bibfield  {journal} {  }\bibfield  {author} {\bibinfo {author} {\bibfnamefont
  {I.}~\bibnamefont {Regev}}, \bibinfo {author} {\bibfnamefont
  {T.}~\bibnamefont {Lookman}}, \ and\ \bibinfo {author} {\bibfnamefont
  {C.}~\bibnamefont {Reichhardt}},\ }\href {\doibase
  10.1103/PhysRevE.88.062401} {\bibfield  {journal} {\bibinfo  {journal} {Phys.
  Rev. E}\ }\textbf {\bibinfo {volume} {88}},\ \bibinfo {pages} {062401}
  (\bibinfo {year} {2013})}\BibitemShut {NoStop}%
\bibitem [{\citenamefont {Fiocco}\ \emph {et~al.}(2014)\citenamefont {Fiocco},
  \citenamefont {Foffi},\ and\ \citenamefont {Sastry}}]{14FFS}%
  \BibitemOpen
  \bibfield  {author} {\bibinfo {author} {\bibfnamefont {D.}~\bibnamefont
  {Fiocco}}, \bibinfo {author} {\bibfnamefont {G.}~\bibnamefont {Foffi}}, \
  and\ \bibinfo {author} {\bibfnamefont {S.}~\bibnamefont {Sastry}},\ }\href
  {\doibase 10.1103/PhysRevLett.112.025702} {\bibfield  {journal} {\bibinfo
  {journal} {Phys. Rev. Lett.}\ }\textbf {\bibinfo {volume} {112}},\ \bibinfo
  {pages} {025702} (\bibinfo {year} {2014})}\BibitemShut {NoStop}%
\bibitem [{\citenamefont {Cundall}\ and\ \citenamefont {Strack}(1979)}]{79CS}%
  \BibitemOpen
  \bibfield  {author} {\bibinfo {author} {\bibfnamefont {P.~A.}\ \bibnamefont
  {Cundall}}\ and\ \bibinfo {author} {\bibfnamefont {O.~D.~L.}\ \bibnamefont
  {Strack}},\ }\href {\doibase 10.1680/geot.1979.29.1.47} {\bibfield  {journal}
  {\bibinfo  {journal} {Géotechnique}\ }\textbf {\bibinfo {volume} {29}},\
  \bibinfo {pages} {47} (\bibinfo {year} {1979})}\BibitemShut {NoStop}%
\bibitem [{\citenamefont {Silbert}\ \emph {et~al.}(2001)\citenamefont
  {Silbert}, \citenamefont {Erta\ifmmode~\mbox{\c{s}}\else \c{s}\fi{}},
  \citenamefont {Grest}, \citenamefont {Halsey}, \citenamefont {Levine},\ and\
  \citenamefont {Plimpton}}]{01SEGHLP}%
  \BibitemOpen
  \bibfield  {author} {\bibinfo {author} {\bibfnamefont {L.~E.}\ \bibnamefont
  {Silbert}}, \bibinfo {author} {\bibfnamefont {D.}~\bibnamefont
  {Erta\ifmmode~\mbox{\c{s}}\else \c{s}\fi{}}}, \bibinfo {author}
  {\bibfnamefont {G.~S.}\ \bibnamefont {Grest}}, \bibinfo {author}
  {\bibfnamefont {T.~C.}\ \bibnamefont {Halsey}}, \bibinfo {author}
  {\bibfnamefont {D.}~\bibnamefont {Levine}}, \ and\ \bibinfo {author}
  {\bibfnamefont {S.~J.}\ \bibnamefont {Plimpton}},\ }\href {\doibase
  10.1103/PhysRevE.64.051302} {\bibfield  {journal} {\bibinfo  {journal} {Phys.
  Rev. E}\ }\textbf {\bibinfo {volume} {64}},\ \bibinfo {pages} {051302}
  (\bibinfo {year} {2001})}\BibitemShut {NoStop}%
\bibitem [{\citenamefont {Akella}\ \emph {et~al.}(2018)\citenamefont {Akella},
  \citenamefont {Bandi}, \citenamefont {Hentschel}, \citenamefont {Procaccia},\
  and\ \citenamefont {Roy}}]{Akella2018}%
  \BibitemOpen
  \bibfield  {author} {\bibinfo {author} {\bibfnamefont {V.~S.}\ \bibnamefont
  {Akella}}, \bibinfo {author} {\bibfnamefont {M.~M.}\ \bibnamefont {Bandi}},
  \bibinfo {author} {\bibfnamefont {H.~G.~E.}\ \bibnamefont {Hentschel}},
  \bibinfo {author} {\bibfnamefont {I.}~\bibnamefont {Procaccia}}, \ and\
  \bibinfo {author} {\bibfnamefont {S.}~\bibnamefont {Roy}},\ }\href@noop {}
  {\bibfield  {journal} {\bibinfo  {journal} {Phys. Rev. E.}\ }\textbf
  {\bibinfo {volume} {98}},\ \bibinfo {pages} {102905} (\bibinfo {year}
  {2018})}\BibitemShut {NoStop}%
\bibitem [{\citenamefont {Miller}\ \emph {et~al.}(1996)\citenamefont {Miller},
  \citenamefont {O'Hern},\ and\ \citenamefont {Behringer}}]{Miller1996}%
  \BibitemOpen
  \bibfield  {author} {\bibinfo {author} {\bibfnamefont {B.}~\bibnamefont
  {Miller}}, \bibinfo {author} {\bibfnamefont {C.~S.}\ \bibnamefont {O'Hern}},
  \ and\ \bibinfo {author} {\bibfnamefont {R.~P.}\ \bibnamefont {Behringer}},\
  }\href@noop {} {\bibfield  {journal} {\bibinfo  {journal} {Phys. Rev. Lett.}\
  }\textbf {\bibinfo {volume} {77}},\ \bibinfo {pages} {3110} (\bibinfo {year}
  {1996})}\BibitemShut {NoStop}%
\bibitem [{\citenamefont {Veje}\ \emph {et~al.}(1999)\citenamefont {Veje},
  \citenamefont {Howell},\ and\ \citenamefont {Behringer}}]{Veje1999}%
  \BibitemOpen
  \bibfield  {author} {\bibinfo {author} {\bibfnamefont {C.~T.}\ \bibnamefont
  {Veje}}, \bibinfo {author} {\bibfnamefont {D.~W.}\ \bibnamefont {Howell}}, \
  and\ \bibinfo {author} {\bibfnamefont {R.~P.}\ \bibnamefont {Behringer}},\
  }\href@noop {} {\bibfield  {journal} {\bibinfo  {journal} {Phys. Rev. E.}\
  }\textbf {\bibinfo {volume} {59}},\ \bibinfo {pages} {739} (\bibinfo {year}
  {1999})}\BibitemShut {NoStop}%
\end{thebibliography}%

\end{document}